\documentstyle[12pt]{article}

\def\Re{\mathop{\rm Re}\nolimits}
\def\title{\bgroup\obeylines\everypar={\hskip\parfillskip}\large
           \bf\vrule height1cm width 0pt\relax}
\def\endtitle{\vskip1sp\egroup}
\def\author#1{\hbox to\textwidth{\hss\vrule height.9cm width0pt\relax #1\hss}}
\def\contauthor#1{\hbox to\textwidth{\hss\vrule width0pt\relax #1\hss}}
\def\moreauthors#1{\hbox to\textwidth{\hss\vrule height.8cm
                   width0pt\relax #1\hss}}
\def\instit{\bgroup\small\it\obeylines\everypar{\hskip\parfillskip}}
\def\endinstit{\vskip1sp\egroup}

\begin{document}

\vglue -4 true cm
\vskip -4 true cm
\begin{center}
{\hfill }{\tt FT-UCM12/92}
\end{center}
\vskip 2 true cm

\begin{title}
The U(1)-Higgs Model:
Critical behaviour in the Confining-Higgs region
\end{title}

\bigskip

\author{J.L.~Alonso, V.~Azcoiti, I.~Campos, J.C.~Ciria, A.~Cruz,}
\contauthor{D.~\'I\~niguez, F.~Lesmes, C.~Piedrafita, A.~Rivero, and
A.~Taranc\'on,}
\begin{instit}
Departamento de F\'{\i}sica Te\'orica, Universidad de Zaragoza,
50009 Zaragoza, Spain,
\end{instit}

\moreauthors{D.~Badoni,}
\begin{instit}
Dipartimento di Fisica, Universit\`a di Roma II, Italy and
INFN  Sezione di Roma -Tor Vergata.
\end{instit}

\moreauthors{ L.~A.~Fern\'andez, A.~Mu\~noz Sudupe, and J.J.~Ruiz-Lorenzo,}
\begin{instit}
Departamento de F\'{\i}sica Te\'orica, Universidad Complutense de Madrid,
28040 Madrid, Spain,
\end{instit}

\moreauthors{A.~Gonz\'alez-Arroyo and P.Mart\'{\i}nez,}
\begin{instit}
Departamento de F\'{\i}sica Te\'orica, Universidad Aut\'onoma de Madrid,
28034 Madrid, Spain,
\end{instit}

\moreauthors{J.~Pech,}
\begin{instit}
Fyzik\'aln\'\i\ \'ustav \v CSAV, Praha, Czechoslovakia,
Dipartimento di Fisica, Universit\'a di Roma I, Italy and
INFN-Sezione di Roma.
\end{instit}

\moreauthors{and}

\moreauthors{P.~T\'ellez,}
\begin{instit}
Servicio de Intrumentaci\'on Electr\'onica, Facultad de Ciencias,
Universidad de Zaragoza, Spain.
\end{instit}

\begin{center}
{September 25, 1992}
\end{center}

\newpage

\begin{abstract}

We study numerically the critical properties of the U(1)-Higgs lattice model,
with fixed Higgs modulus, in the region of small gauge coupling where the
Higgs and Confining phases merge. We find evidence of a first order transition
line that ends in a second order point. By means of a rotation in parameter
space we introduce thermodynamic magnitudes and critical exponents in close
resemblance with simple models that show analogous critical behaviour. The
measured data allow us to fit the critical exponents finding values in
agreement with the mean field prediction. The location of the critical point
and the slope of the first order line are accurately given.

\end{abstract}

\newpage

\section{Introduction}

The nonperturbative formulation of four dimensional field theories is a very
debated subject.  At present, only asymptotically free theories can be
rigorously constructed (See \cite{CALL} for a review).

On the other hand, there is evidence \cite{AIZEN,FREED} that a self-coupled
($\lambda \varphi^4$) four dimensional scalar field theory is trivial, that
is, it describes a free theory after ---nonperturbative--- renormalization.

The question of nontriviality of scalar fields coupled to gauge fields is,
however, not so clear \cite{CALLPETR}. In the last few years a considerable
effort on the understanding of this problem has been carried out.

In this work we study a four dimensional theory with a continuous symmetry
group: the fixed module U(1)-Higgs theory. Although it does not represent a
limit of the SU(2)$\times$U(1) theory (it lacks a global SU(2) symmetry), we
expect that many of the results obtained here may be useful for more complex
models.

The phase diagram of our model is represented in Fig.~1 where it can be seen
that there are three phases: Confining, Higgs and  Coulomb (strictly speaking
only two since the first and second are analytically connected).

We have focused our attention in the line that separates the first two phases.
We found that it corresponds to a first order phase transition. At the
end-point of the line we have observed a clear critical (second order)
behaviour, where it is possible to define a continuum limit. We study the
critical exponents at this point, as they are useful to discover the
properties of the continuum theory, in particular whether the theory is
trivial or interacting.

We study the model with the parameter $\lambda=\infty$ (see (\ref{ACT1})),
which
is equivalent to fix the modulus of the Higgs field $|\Phi|=1$. It is
generally assumed that the fixed modulus theory belongs to the same
Universality Class as the full ($\lambda$ finite) one.

In order to find the critical exponents we compute the evolution of several
thermodynamic quantities for different coupling and several lattice sizes.

We have performed Monte Carlo simulations on lattices ranging from $6^4$ to
$24^4$. The results presented here amount to 8 months of CPU of a custom 64
INMOS T805 transputer machine with a performance of 100 Mflops.

In section 2 and 3 we formulate the model and describe some of its known
properties. Scaling relations and critical exponents near the critical point
are also introduced. In section 4 we define the observables that will be
measured in the simulation. The numerical method is described in section 5.
Finally, the results are shown in section 6. We include in the appendix the
description of the dedicated multiprocessor machine designed and constructed
by our group.

\section{Formulation}

The action for a self-coupled scalar field with a local U(1) gauge symmetry
(U(1)-Higgs model) in a lattice can be written as

\begin{equation}
S=-\beta\sum_{{\bf r},\mu<\nu} \Re U_{{\bf r},\mu\nu}
  -\kappa\sum_{{\bf r},\mu} \Re\bar\Phi_{\bf r}U_{{\bf r},\mu}\Phi_{{\bf
r}+\mu}
  +\lambda\sum_{\bf r}(|\Phi_{\bf r}|^2-1)^2
  +4\kappa\sum_{\bf r}|\Phi_{\bf r}|^2,
\label{ACT1}
\end{equation}
where ${\bf r}$ is the four dimensional lattice site, the Greek indices
$\mu,\nu\in\{1,2,3,4\}$ represent the four lattice directions and $\Phi_{\bf
r}$ is the value of the complex scalar field at ${\bf r}$, $U_{{\bf r},\mu}$
is the gauge (U(1)) field at the link in the $\mu$ direction beginning at
${\bf r}$, and $U_{{\bf r},\mu\nu}$ is the plaquette defined by the site ${\bf
r}$ and the directions $\mu$ and $\nu$.

It is usually assumed that the action (\ref{ACT1}) for finite $\lambda$ belongs
to the same universality class that the one in the $\lambda\to\infty$ limit.
Taking this limit, we can fix the modulus of the scalar field, and the action
becomes (up to an additive constant)

\begin{equation}
S=-\beta\sum_{{\bf r},\mu<\nu} \Re U_{{\bf r},\mu\nu}
-\kappa\sum_{{\bf r},\mu} \Re\bar\Phi_{\bf r}U_{{\bf r},\mu}\Phi_{{\bf r}+\mu}
\label{ACT2}
\end{equation}

In this way both the gauge and the scalar field belong to the U(1) group. We
will limit ourselves in this work to study this two--parameter model.

\section{Critical behaviour}

\subsection{Description of the parameter space}

Let us briefly describe the different limits of the action when the coupling
parameters $\beta ,\kappa$ take the extreme values 0 or $\infty$.

\subsubsection{$\beta=\infty$}

In this case the gauge fields are frozen out and the remaining model is:
\begin{equation}
S=-\kappa\sum_{{\bf r},\mu} \Re\bar\Phi_{\bf r} \Phi_{{\bf r}+\mu}
\end{equation}
This action has a global O(2) symmetry (or U(1)). The O(N) model in four
dimensions shows a continuous transition between a disorder phase, with
explicit O(N) symmetry at low $\kappa$ and a ordered one, at high $\kappa$,
where the O(N) symmetry breaks down to a O(N-1) symmetry . Due to the
Goldstone theorem this phase has N-1 massless Goldstone bosons  (spin waves).
This is a gaussian second order phase transition \cite{O4MODEL}.

We can estimate $\kappa_c$ using the mean field approximation (MFA), the result
is \cite{LARKIN}:
\begin{equation}
\kappa_{c,{\rm MFA}}=\frac{N}{2 q}
\end{equation}
where $q$ is the coordination number, $2 d$ in our case ($d$-dimensional
square lattice). This is an approximation from below since  MFA neglects the
fluctuations and overestimates the interactions.

An upper limit is \cite{SHROCK}:
\begin{equation}
\kappa_c\le \frac{N I(d)}{2}
\end{equation}
where:
\begin{equation}
I(d)=\int_{-\pi}^{\pi} \frac{dk_1...dk_d}{(2\pi)^d} \frac{1}{\sum_{j=1}^d
(1-\cos(k_j))}
\end{equation}

If we take $N=2$ and $d=4$ we get:
\begin{equation}
0.125\le\kappa_c\le 0.311
\end{equation}

\subsubsection{$\beta=0$}

This situation corresponds to a spin model with annealed bond disorder
\cite{SHROCK} (the disorder is in thermodynamic equilibrium with the Higgs
field). The action is:

\begin{equation}
S=-\kappa\sum_{{\bf r},\mu} \Re\bar\Phi_{\bf r}U_{{\bf r},\mu}
\Phi_{{\bf r}+\mu}
\end{equation}

There is not a phase transition in this case, for the following reason: it can
be made a gauge transformation which maps the Higgs field into a constant
because the Higgs field lives in the U(1) group. The functional integration
over the Higgs is trivial and the U(1) sector is a one-link separable theory
(without interaction).

\subsubsection{$\kappa=0$}

The remaining model is $U(1)$ pure gauge. This model shows a transition
between a maximally disordered Confining phase, at low $\beta$, and a Coulomb
phase (it has free photons and coulomb potential between static charges) at
high $\beta$. The transition, located at $\beta\approx 1$, is found to be
first order \cite{U1GAUGE}.

Applying a theorem by Shrock \cite{SHROCK}, the full theory at low $\kappa$
can be written as a pure U(1)  theory with a shifted coupling, where the shift
is proportional to $\kappa^4$. Thus we can extend the $\kappa=0$ transition to
a small $\kappa$.

\subsubsection{$\kappa=\infty$}

It is always possible to fix the unitary gauge, so that the Higgs fields
disappears and the action (\ref {ACT2}) becomes

\begin{equation}
S=-\beta\sum_{{\bf r},\mu<\nu} \Re U_{{\bf r},\mu\nu}
 -\kappa\sum_{{\bf r},\mu} \Re U_{{\bf r},\mu}
\label{ACTUP}
\end{equation}

Now in the $\kappa=\infty$ limit, the second term in the action makes that
only configurations that satisfy $\Re U_{{\bf r},\mu}=1$ have non vanishing
probability. Then, $S$ becomes trivial, and there is no transition in this
case. The situation changes when charged ($q>1$) Higgs fields are considered
\cite{ALFONSO}. Then the limit $\kappa=\infty$ corresponds to a non trivial
$Z_q$ gauge theory.

\subsubsection{Interior of the parameter space}

The phase diagram of the model was studied on the pioneering work of Fradkin
and Shenker \cite{FRADKIN}, and is plotted in the figure 1. Further studies
can be found in references \cite{O4MODEL,U1GAUGE,VICENTE}.

On the point ``C" there is coexistence of Confining, Coulomb and Higgs phases
(triple point).

If we consider $\lambda \ne \infty$ the main modification is that the end
point ``D" tends, as $\lambda$ decreases, to the $\beta=0$ axis and finally
cuts it.

The transition line ``B-C" is a line of second order transition, the vertical
line ``A-C" is of first order \cite{VICENTE} and we will show that the line
``D-C" is, again, of first order. The end point (``D") is according to our
study a second order point.

\subsection{Critical behaviour in related systems}

Regarding the ``C-D" line, the structure of our model is similar to that of a
wide variety of systems, like, for instance, the magnetic Ising model or the
liquid-vapour transition.

The Ising model with nearest neighbour interaction has an action
\begin{equation}
S=-J \sum_{<ij>} S_i S_j -h \sum_i S_i
\end{equation}
it shows a first order phase transition line ($h=0, J>J_c$) that finishes in a
second order critical point at $J_c$.

The position of the end point may be determined studying the behaviour of the
magnetization over the first order --straight--- line. It is found that

\begin{equation}
M\sim (J-J_c)^{\beta},\quad J>J_c
\end{equation}
The exponents $\alpha$ and $\gamma$ are defined respectively from the critical
behaviour of the specific heat and susceptibility:
\begin{eqnarray}
C\sim |J-J_c|^{-\alpha} \\
\chi\sim |J-J_c|^{-\gamma}
\end{eqnarray}
which hold the scaling relation \cite{STANLEY}
\begin{equation}
\alpha+2\beta+\gamma=2\label{SCALREL}
\end{equation}

Another analogous system is the liquid-vapour transition. The lack of a
symmetry implies that neither the straightness of the critical line nor its
exact location  ($h=0$ in the Ising case) are given. In this system the
critical exponents ($\alpha,\beta,\gamma$) have been experimentally measured
\cite{MA} and the relation (\ref{SCALREL}) checked.

In our model we have not an explicit symmetry and so, the exact position and
shape of the transition line, must be numerically computed. However we can
define the critical exponents as in the previous models.

\section{Observables}

We define the (normalized) plaquette and link energies as

\begin{eqnarray}
E_{P}=\frac{1}{6V}\sum_{{\bf r},\mu<\nu}\Re U_{{\bf r},\mu\nu}\\
E_{L}=\frac{1}{4V}\sum_{{\bf r},\mu}
		\Re\bar\Phi_{\bf r}U_{{\bf r},\mu}\Phi_{{\bf r}+\mu}
\end{eqnarray}
where $V=L^4$ is the volume of the lattice. In terms of the above energies,
the action can be rewritten as:
\begin{equation}
-S=\beta E_{P} 6V+\kappa E_{L}4V
\end{equation}
Both $E_P$ and $E_L$ lie in the $[-1,1]$ interval. With our definition,
$E_P\to 1$ when $\beta\to\infty$ and $E_L\to 1$ when $\kappa\to\infty$.
$<E_P>=<E_L>=0$ at $\beta=\kappa=0$.

Let us write the partition function as
\begin{equation}
{\cal Z}(\beta,\kappa)=\int [dU] [d\Phi] e^{-S}.
\end{equation}
It is useful to introduce the parameter $\kappa'\equiv 2\kappa/3$ to
symmetrize the action. In this way the  fluctuation matrix (or connected
correlation) can be written as
\begin{equation}
F_{ij}\equiv<E_i E_j>-<E_i><E_j>=\frac{1}{(6 V)^2}
\frac{\partial^2 \log {\cal Z}(x_i,x_j)}{\partial x_i \partial x_j}
\end{equation}
where
\begin{equation}
x_1=\beta,\ x_2=\kappa',\ E_1=E_P,\ E_2=E_L
\end{equation}

At a given point $(\beta_0,\kappa'_0)$, the $F$ matrix can be diagonalized. We
shall call $\lambda_{\rm max}$ and $\lambda_{\rm min}$ to their maximum and
minimum eigenvalues respectively.

We can perform a rotation of angle $\theta$ in such a way that in the new
coordinates

\begin{eqnarray}
c_\perp\equiv\hphantom{-}\beta \cos\theta + \kappa' \sin\theta\\
c_\parallel\equiv      - \beta \sin\theta + \kappa' \cos\theta
\end{eqnarray}
the fluctuation matrix is diagonal.

Consequently the operators
\begin{eqnarray}
E_\perp\equiv\hphantom{-} E_1 \cos\theta+E_2 \sin\theta\\
E_\parallel\equiv -E_1 \sin\theta+ E_2\cos\theta
\end{eqnarray}
are uncorrelated. In terms of the new quantities the action
can be written as
\begin{equation}
-S= (c_\perp E_\perp + c_\parallel E_\parallel) 6V
\end{equation}

Assuming that the point ``D" is second order, we expect divergences in some
magnitudes. Using the Ising model analogy discussed in section (3.2) we can
write the following formulae for the previously defined thermodynamic
quantities:
\begin{equation}
\Delta E_\perp(c_\parallel)
 \equiv\left.\frac{\partial f}{\partial c_\perp}\right|_{c_\perp=a^+}-
 \left.\frac{\partial f}{\partial c_\perp}\right|_{c_\perp=a^-}\sim
(c_\parallel-c_\parallel^c)^\beta, \quad c_\parallel<c_\parallel^c\label{BETA}
\end{equation}

\begin{equation}
\chi(c_\parallel)\equiv\left.\frac{\partial^2 f}
{\partial c_\perp^2}\right|_{c_\perp=a} =(6V)\lambda_{\rm max} \sim
|c_\parallel-c_\parallel^c|^{-\gamma }
\end{equation}

\begin{equation}
C(c_\parallel)\equiv\left.\frac{\partial^2 f}{\partial
c_\parallel^2}\right|_{c_\perp=a} = (6V) \lambda_{\rm min}
\sim |c_\parallel-c_\parallel^c|^{-\alpha}
\end{equation}
where $f\equiv \frac{1}{6V}\log {\cal Z}$ is the intensive free energy,
$\chi$
is the susceptibility and $C$ is the specific heat. We have call $a$ to the
value of the $c_\perp$ parameter on the first order line that, as we shall see
below, is almost independent of $c_\parallel$ in the interesting region
(neighbourhood of ``D").

The critical law for  $\Delta E_\perp$ is analogous to that for the Ising
magnetization. We denote as $\Delta E_\perp$ the difference between $E_\perp$
in the Confining and Higgs phases in the $c_\perp\to a$ limit.

Therefore, we expect the three critical exponents defined above to follow the
scaling relation (\ref{SCALREL}).

\section{The numerical method}

We have used the subgroup $Z_N$ with $N=1024$ as an approximation of the gauge
group $U(1)$ since,  in the region of interest, the fluctuations of the
variables are large (typically of the order of 1 radian),  and the phase
transition associated with the discrete group is safely far away.

The updating algorithm is an adaptive step size Metropolis, with an acceptance
rate of more than a sixty percent.

\subsection{Parallelization}

We have used lattice sizes $6^4, 8^4, 12^4, 16^4$ and $24^4$ implemented on a
transputer (IMS-T805) \cite{TRANSPUTERS} machine of 64 processors with a
$8\times 8$ topology.  For the smaller lattices, we have also used transputer
boards of 8 and 3 processors with $4\times 2$, $8\times 1$ and $3\times 1$
topologies.

The parallelization strategy is straightforward: divide the lattice among the
processors, so that each of them accounts for a smaller sublattice that can be
updated in parallel with the other transputers.  One of the problems is that,
being a nearest neighbour interaction, the update cannot be entirely made
inside each processor: the sites and links in the border of the sublattice
must know some of the variables in the neighbouring transputer.

A critical choice in the parallelization is the way in which neighbouring
processors exchange  the necessary information during the update. The simpler
method could be the transmission, by means of a parallel process, of the
variables that are required in a given step of the calculation. In addition of
a minor problem of synchronization, transmitting single variables is poorly
efficient regarding the link bandwidth and also, it means a considerable
overhead due to the frequent start and end of needed parallel processes.

For this reason we have added to the sublattice in each transputer the rows
and columns of the links and sites needed to make the update entirely inside.
For instance, in the $16^4$ lattice on the 64-processor machine with a
$8\times 8$ topology, each transputer holds a sublattice of
$4\times 4\times 16^2$ lattice and not a $2\times 2\times 16^2$ lattice. The
update is performed almost synchronized in all processors. Once it is
finished, the borders are transmitted to the neighbouring processors. To
guarantee the independence of the variables currently updating, we perform a
checkered update.

We have obtained in this way a parallel efficiency close to a 95\%.

\subsection{Spectral Density Method}

The precise location of the critical values of the parameters may be a
difficult task because Monte Carlo methods provide information about the
thermodynamic quantities only at particular values of the couplings.  The
approach that we use here to locate them is based on histograms and is known
as the Spectral Density Method \cite{FALCIONI}.

The generalization of the method to a two--dimensional parameter space is
straightforward. Considering that, we perform a Monte Carlo simulation at a
particular point of the parameter space $(\beta,\kappa)$, and we compute the
histogram $H(E_P,E_L)$ as an approximation to the density of states. The
probability of finding a configuration of plaquette energy $E_P$ and link
energy $E_L$ at a different point $(\beta_1,\kappa_1)$ can be calculated as

\begin{equation}
P^{(\beta_1,\kappa_1)}(E_P,E_L)=
     \frac{H(E_P,E_L) e^{(\beta_1-\beta) 6VE_P+(\kappa_1-\kappa) 4VE_L}}
    {\int dE'_PdE'_L H(E'_P,E'_L) e^{(\beta_1-\beta) 6VE'_P+(\kappa_1-\kappa)
					 4VE'_L}}.\label{PROB}
\end{equation}

Let us discuss the range of applicability of the Spectral Density Method. Let
$\sigma_P,\sigma_L$ be the widths of the measured histogram in the $E_P,E_L$
directions respectively. It may be easily seen from the previous equation that
the ranges are $\Delta\beta\sim 1/(6V\sigma_P), \Delta\kappa\sim
1/(4V\sigma_L)$.

Although near the transition line $\sigma_P$ (or analogously $\sigma_L$) is
large, the application of (\ref{PROB}) at fixed $\kappa$ is very useful to
find the $\beta$-value where the fluctuation of the energy has a maximum
---apparent critical point--- and, eventually, to adjust the parameters for a
new simulation.

We can also move simultaneously in both directions. The minimum eigenvalue of
the fluctuation matrix corresponds to an eigenvector parallel to the transition
line. This means that the range of applicability of the Spectral Density Method
in the $c_\parallel$ direction is large.

We can go one step further using data from simulations at different points of
the parameter space, using a two--dimensional generalization of the
multihistogram method proposed in \cite{FERRENBERG}, which gathers all the
information for a given lattice size. This method has been used for many
observables and the results plotted as a smooth line in the figures, with the
points corresponding to single simulations. However, in order to obtain a
safer estimation of the errors, we only use the single points to do the fits
to a critical power law.

Usually, the Spectral Density Method is very useful to find the value of the
coupling where some observable has a maximum, as well as to obtain an accurate
value for this maximum. In this work we have used it extensively to locate the
transition line ``C-D'' at a fixed $c_\parallel$, looking for the maximum in
the fluctuation of $E_\perp$. However, in order to find the end point ``D'' we
have to move in the $c_\parallel$ direction, in doing so we do not find a
maximum for any simple observable, and the usefulness of the method is a great
extent lost.

\subsection{Measurements}

At every simulation point we store the plaquette and link energies to
construct the histograms. We compute the two--dimensional histogram in the
$E_P$-$E_L$ plane. As an example, we show in figure 2 some contour plots of
the energies histogram.

In principle we can move in the $(\beta,\kappa)$ plane in whatever direction
using the spectral density method. In figure 3 we plot the mean value of
$E_{\perp}$  as a function of the couplings obtained with the multihistogram
method.

Nevertheless, most of the results presented in this work, have been obtained
studying one dimensional histograms. To this end, we rotate the parameter
space in order to discretize the energies $E_{\perp}$.  To simplify the
computations we have chosen a fixed rotation angle for all lattice sizes and
all the parameter space points. Notice that the error in the determination of
the angle could only mean second order corrections for most of the quantities
of interest. The only exception is the minimum eigenvalue $\lambda_{\rm min}$
that is computed diagonalizing the fluctuation matrix in every simulation.

We usually discretize the energy interval into one hundred subintervals. From
the (one dimensional) histogram (figure 4, upper side), we compute $E_\perp$
and $\partial E_\perp/\partial c_\perp$ in the neighbourhood of the simulation
point, determining the best approximation to the critical point by looking at
the maximum of the derivative. In figure 5 we show an example of this method.
The validity range is estimated from the fluctuation in the energy.

A very important quantity to be computed is the latent heat, that is, the
difference between the values of $E_\perp$ on both sides of the first order
line. In a finite lattice this limit is not well defined; we take as its
definition the distance between the two minima of the effective potential at
the first order line (that is, the distance between the two maxima of the
histogram). Unfortunately, this procedure needs  a precise estimation of the
local maxima of a noisy function. To reduce the statistical error, we smooth
the histogram in the region near each maximum with a cubic spline, measuring
the distance between the smoothed functions (see figure 4, lower side).

The computation of statistical errors has been carried out using the jack-knife
method. We perform a primary determination of the correlation in Monte Carlo
time and construct statistically independent bins. The number of iterations
performed in the larger lattices for $\kappa\in[0.52,0.54]$ has been (in
thousands of Monte Carlo Sweeps)
\begin{equation}
\begin{array}{ll}
L=16,\ \kappa=0.52:\quad   &  200\\
L=16,\ \kappa=0.525:\quad  &  500\\
L=16,\ \kappa=0.5275:\quad &  600\\
L=16,\ \kappa=0.53:\quad   &  800\\
L=16,\ \kappa=0.54:\quad   &  200\\
L=24,\ \kappa=0.52:\quad   &  100\\
L=24,\ \kappa=0.5275:\quad &  800\\
L=24,\ \kappa=0.53:\quad   &  200\\
\end{array}
\end{equation}
so that, in the neighbourhood of the critical point, we can use bins of around
a hundred thousand of MC sweeps.

\section{Results}

In this section we report our results classified according to every observable
computed. In all cases we have used the coordinate transformation to the
$(c_\perp,c_\parallel)$ plane.

For the sake of simplicity regarding the figures we fix from the beginning
\begin{equation}
\theta  \equiv 0.96 \label{THETA}
\end{equation}
which is within a 1\% our estimation for the angle of the fluctuation matrix
defined above. We will discuss in every case the effects of the particular
selection of this quantity.

We have always run in points of the parameter space over the first order line
for $c_\parallel<c_\parallel^c$ and on the prolongation (dotted line in figure
1) for $c_\parallel>c_\parallel^c$.

Since there is not an explicit symmetry, as in the liquid vapour transition,
the straightness of the transition line is not implied, and then the angle of
the eigenvector of the fluctuation matrix does not have to take the same value
as the one of the first order line. However we have found numerically that,
over the line, $c_\perp$ is practically constant; it changes less than
$0.01\%$ for $\kappa \in[0.52,0.54]$ ($c_\parallel\in[-0.501,-0.478]$).

\subsection{Latent Heat}

As we saw in (\ref{BETA}) the critical exponent $\beta$ is related with the
behaviour of $\Delta E_\perp$ over the critical line.

At the first order line there is a discontinuity in $E_\perp$ and the gap
between its values on both sides, is, in the thermodynamic limit,  the latent
heat. In a finite lattice it is not easy to measure this gap since the
discontinuity is rounded. The method used in this work is to compute the gap
from the histogram in the quantity $E_\perp$. More precisely we compute the
distance between the two maxima of the histogram. If the height of both is not
the same we shift (with the Spectral Density Method) the histogram the
necessary amount in $c_\perp$.

Since $E_\parallel$ is continuous, it is equivalent, regarding the critical
behaviour, to make the analysis in terms of $E_\perp$ or almost any linear
combination of $E_\perp$ and $E_\parallel$, with the only requirement of a
nonvanishing coefficient in $E_\perp$. In \cite{LETTER} we used the latent
heat for the plaquette energy. We expect a ---slightly--- better measurement
when choosing the optimum combination of $E_P$ and $E_L$ which is $E_\perp$.

For the same reason, a small error in the determination of the rotation angle
$\theta$ is unimportant. We remark that this error affects only quadratically
so that it is completely negligible.

In figure 6 we show our measurements for $\Delta E_\perp$ for several
couplings and lattice sizes as a function of $c_\parallel$.

Although the critical behaviour is very clear, a precise determination of the
critical exponent ($\beta$) is very difficult.

Equation (\ref{BETA}) is only followed strictly in the thermodynamic limit. In
a finite lattice, we should find deviations from the functional form
$E_\perp=A(c_\parallel-c_\parallel^c)^\beta$, however, it is expected that the
main deviation can be considered as a finite size dependence on the parameters
$\{A, c^c_\parallel,\beta\}$. In particular we expect the strongest dependence
on the parameter $c^c_\parallel$ (see section below on Finite Size Scaling).
The procedure we use to compute the parameters is the following: we fit the
whole data to the function (\ref{BETA}) with $\beta$ independent of $L$,
allowing a finite size dependence for $c^c_\parallel$ and $A$. Successively we
discard the data from the smaller lattices to check the asymptotic behaviour.
Our results are

\begin{equation}
\begin{tabular}{ll}
$L=6,8,12,16$ & $\beta=0.53(7)$\\
$L=8,12,16$ & $\beta=0.53(7)$\\
$L=12,16$ & $\beta=0.55(9)$ \\
$L=16$   & $\beta=0.50(11)$
\end{tabular}
\end{equation}

Although the statistical errors do not allow us to observe a monotonous
evolution to the thermodynamic limit, the previous results give a strong
evidence in favour of the classical value $\beta=1/2$.

In reference \cite{LETTER} we compute the exponent $\beta$ using only the data
from the plaquette energy. Although the simulations are essentially the same,
some minor variations are found since the observables are not completely
correlated. In \cite{LETTER} we found $\beta=0.54(6)$ for $L=12,16$ and
$\beta=0.47(9)$ for $L=16$.

In figure 7 we plot the latent heat squared as a function of the parameter
$c_\parallel$ for the points near the critical one. The linear behaviour seems
to be in agreement with the data.

\subsection{Maximum eigenvalue}

An alternative way to compute the exponent $\beta$ is measuring the
fluctuation of the energy. The fluctuation of $E_\perp$ is what we previously
called $\lambda_{\rm max}$. In the limit of a histogram with infinitely narrow
peaks, $\lambda_{\rm max}=(\Delta E_\perp)^2/4$. We point out that measuring
$\lambda_{\rm max}$ instead of $\Delta E_\perp$ is similar to measure the
square of the magnetization (in a magnetic system) to avoid the cancellation of
the magnetization due to tunneling effects.

In practice it is not necessary to diagonalize the fluctuation matrix at each
point since the variations of the angle $\theta$ only affect quadratically,
and mainly as a global multiplicative constant.

In figure 8 we show the evolution of $\lambda_{\rm max}$ as a function of
$c_\parallel$ for several lattice sizes. We see a window in the larger
lattices where the behaviour is almost linear according to $\lambda_{\rm
max}=A(c_\parallel-c_\parallel^c)^{2 \beta}$ with $\beta$ near $1/2$.

We remark that the statistical error in the measure of $\lambda_{\rm max}$ is
much smaller than the one for the latent heat (compare figures 7 and 8).
Unfortunately, the deviation from the square of the latent heat due to finite
size effects is large, so it is not easy to obtain a precise estimation of the
exponent $\beta$ and of its error.

\subsection{Susceptibility}

The susceptibility $\chi=\partial E_\perp/\partial c_\perp$ is related to the
maximum eigenvalue of the fluctuation matrix in the absence of phase
coexistence and there its measurement is straightforward: $\chi=6V\lambda_{\rm
max}$. When $c_\parallel<c_\parallel^c$ we would have to measure $\chi$ at
$c_\perp$ near the transition line and, after that, take the limit from one
side of the transition. Alternatively, and this is the method we use, we can
divide the histogram in $E_\perp$ in two halves and measure the fluctuation on
each half. In the thermodynamic limit the results are equivalent. However, for
finite lattices near the critical point the overlap between both peaks is big
and the measure of $\chi$ cannot be very precise.

In figure 9 we show our results for $6V\lambda_{\rm max}$ which can be called
susceptibility only for $c_\parallel>c_\parallel^c$.

The finite size effects do not allow us to use directly the points in the
simulated lattices for fitting to the critical power law (see figure 9). To
compute $\gamma$ we first take the thermodynamic limit and then fit the
asymptotic values. We must point out that this indirect process reduces the
objectivity in the determination of the error. Our results are
\begin{equation}
\gamma=1.13(17)
\label{gamma}
\end{equation}
according with the Mean Field prediction $\gamma=1$.

In figure 10 we plot the results for $\chi$ in the larger lattices, including
also the values for $c_\parallel<c_\parallel^c$ obtained using the division
method discussed above. We can obtain a more precise value computing the
dispersion excluding the region of the histogram between peaks: we compute the
dispersion in the first phase integrating over the left part of the first
peak, and in the second, integrating over the right part of the second peak.
The accumulated error is difficult to compute, but we can obtain a crude
estimation comparing the results from both peaks (see figure 10).

We stress that with the hypothesis of the same exponents on both sides of the
transition ($\gamma=\gamma'$) we obtain again (after excluding the points
closer to the critical point which show great finite size effects) a value for
$\gamma$ near 1.

Finally we remark that the divergence of the susceptibility when approaching
the thermodynamic limit is a clear evidence of a second order behaviour at the
end point ``D''.

\subsection{Specific Heat}

The small eigenvalue of the fluctuation matrix is related with the energy
$E_\parallel$:
\begin{equation}
\lambda_{\rm min}=\frac{1}{6V}\frac{\partial E_\parallel}{\partial c_\parallel}
\end{equation}
so that, in our analogy with the Ising model, we can call it Specific Heat.
Since $\partial E_\parallel/\partial c_\parallel\approx
A|c_\parallel-c_\parallel^c|^{-\alpha}$, $V\lambda_{\rm min}$ should present a
divergence at the critical point if $\alpha>0$. If $\alpha= 0$ the divergence
should be logarithmic and so, hard to observe.

To determine the small eigenvalue of a matrix with a much larger one  is not
an easy task. However we have obtained quite precise determinations.  Notice
that the minimum eigenvalue is just the width of the two--dimensional histogram
(see figure 2) which can be clearly distinguished from its length.

In figure 11 we plot the minimum eigenvalue (times $V$) as a function of
$c_\parallel$. The absence of a divergence at the critical point practically
excludes the possibility of a positive value for $\alpha$.

In a strict sense, when there is phase coexistence we would have to compute
the minimal eigenvalue for both phases. For large lattices the $E_P-E_L$
histogram becomes very narrow, and the difference in the angle at each phase
makes the minimum eigenvalue for the whole histogram to grow (see figure 11).
Nevertheless, this does not change our conclusions about the behaviour at the
critical point.

\subsection{Finite Size Scaling}

The critical exponent $\nu$ has been computed with a Finite Size Scaling
Analysis. We have study the shift in the apparent critical point as a function
of the lattice size. We expect that the shift will follow the law
\begin{equation}
\Delta c^c_\parallel(L) \sim L^{-\frac{1}{\nu}}
\end{equation}

In figure 12 we plot the values (squared) obtained fitting the latent heat
data for each lattice size fixing the value of $\beta$. A least squares three
parameter fit gives
\begin{equation}
\nu=0.52(4)
\label{PRIMERNU}
\end{equation}
The value (\ref{PRIMERNU}) is almost insensitive to the value of $\beta$ used
in the extrapolation. For values of $\beta$ in the range $[0.4,0.6]$ the
resulting $\nu$ varies by less than $0.01$.

In reference \cite{LETTER} using just the plaquette energy data, we obtained
the value $\nu=0.47(4)$.

\section{Conclusions}

We have found a second order point where critical exponents
$\alpha,\beta,\gamma$ may be defined in close analogy with ferromagnetic spin
systems. The location of this point has been accurately measured:

\begin{equation}
\begin{tabular}{rl}
$c^c_\perp$&$=0.77391(2)$\\
$c^c_\parallel$&$=-0.494(1)$\\
\end{tabular}
\end{equation}
it corresponds to $\beta^c=0.8485(8),\ \kappa^c=0.5260(9)$. Although we have
fixed in (\ref{THETA}) the rotation angle $\theta$ that defines $c_\perp$ and
$c_\parallel$, our best estimation of the angle that diagonalizes the energy
fluctuation matrix at the critical point is $\theta^c=0.963(3)$.

The thermodynamic limit extrapolations of the latent heat data corresponding
to simulations performed below $c_{\parallel}^c$ clearly give nonvanishing
values, that show the first order nature of the critical line between the
Higgs and the Confining phase.

Fitting the thermodynamic limit extrapolations of the inverse susceptibility we
have seen that it approaches zero at $c_{\parallel}^c$, giving us a strong
evidence of its second order behaviour. This result has also been confirmed
with extrapolations from the region below $c^c_\parallel$.

Our estimations of the critical exponents are compatible with the mean field
results $\alpha=0, \beta=1/2, \gamma=1, \nu=1/2$.

\bigskip
{\large Acknowledgments}
\smallskip

We have benefited from conversations with M. Asorey and A. Sokal. This work
has been supported by the {\it Diputaci\'on General de Arag\'on} (P IT- 2/89)
and the {\it Programa Nacional de Tecnolog\'{\i}as de la Informaci\'on y las
Comunicaciones} (CICyT TIC1161/90-E y TIC91-0963-C02). We also acknowledge
partial financial support from CICyT, projects AEN90-0034, -0272 and -0030.

\section{Appendix}
The 64-processors machine that we have call Reconfigurable Transputer Network
(RTN), has been entirely designed and build inside our group. It provides us
with a power of 100 (sustained) Mflops at a low cost. In the actual
configuration it holds eight boards with eight T800 each, plus a controller
board with one transputer.  RTN interfaces with a host computer, a PC in the
actual configuration, via another board (root board).  The eight transputer
boards have also a {\it cross-link} C004 (programmable switch) that allows to
interconnect them. Each transputer has 1 Mb of memory (may have up to 4Mb).
They are connected inside the board forming a ring.  One of the two remaining
links of each transputer is connected to the C004 and the other free link is
connected to the C004 of the next board.  In this way we may have a torus
($8\times 8$) topology for the 64 transputers.  Nevertheless it is possible
through the C004s to attain different topologies. We could divide, for
instance, RTN in eight identical machines dedicated to different problems.

The purpose of the controller board, which has a transputer and a cross-link ,
is to boot the code to each transputer of RTN, close the torus, wait for the
end of the calculation, open the torus, read the results and repeat the
process.  The way to know when the calculation is done and when to open the
torus is through the {\it event} pin of one of the transputer in each board.

The root board, as we said, interfaces with the host computer and is
responsible for sending the code and input to the controller board, it also
reads the output from the controller and writes it on the host devices (disk,
terminal, etc.).

\newpage
{\large Figure captions.}

\begin{enumerate}

% PHASEDIAG
\item Scheme of the phase diagram of the fixed module U(1)-Higgs model.

% HISTOG2D
\item Contour plots of two--dimensional ($E_P$-$E_L$) energy histograms for
$L=8$ (upper part) and $L=12$ (lower part). The left side ones correspond
to the point $(\beta,\kappa)=(0.854,0.52)$ and those on the right to
$(\beta,\kappa)=(0.835,0.54)$. In all cases the most external contour is
plotted
at a value of a 10\% of the maximum.

% FS2D
\item $E_\perp$ near the critical point in a $L=8$ lattice using a
multihistogram method.

% HIST-LAT
\item Example of one--dimensional histogram (above).
Example of the measure of the latent heat with a cubic spline fit (below).
The data correspond to 500000 sweeps on a $L=16$ lattice for
$(\beta,\kappa)=(0.8495,0.525)$.

% E(C)-D
\item $E_\perp$ (above) and
 $\partial E_\perp/\partial{c_\perp}$ (below) using the Spectral Density
Me\-thod
obtained from a simulation with
$L=12$, $\beta=0.84945$, $\kappa=0.525$, $500000$ MC sweeps. The filled
circle is plotted at the $c_\parallel$ of the simulation.

% LAT-CPAR
\item Jump of $E_\perp$ across the first order
line.
The continuous lines are obtained from a power law fit.

% LAT2CPAR
\item Square of the Latent Heat as function of $c_\parallel$. The straight
lines
are minimum squares fits. Symbols as in figure 6.

% CESPCPAR
\item Maximum eigenvalue of the energy fluctuation matrix
against $c_\parallel$. The continuous lines are obtained with a
multihistogram method.

% CHICPAR
\item Inverse of the susceptibility (when there is no phase coexistence)
as a function of $c_\parallel$. The continuous lines are obtained with a
multihistogram method.

% CHICPAR2
\item Inverse of the susceptibility on both sides of
$c_\parallel^c$ . Only represented for $L=16$ (circles) and $L=24$ (triangles).

% SMALLEIG
\item Minimum eigenvalue of the energy fluctuation matrix against
$c_\parallel$.

% CCRITL
\item $c_\parallel^c(L)$ as a function of $1/L$

\end{enumerate}

\bigskip


\newpage
\begin{thebibliography}{99}


\bibitem{CALL}
D.J.E. Callaway {\sl Phys. Rep.} Vol {\bf 165},5 (1988).

\bibitem{AIZEN}
M. Aizenman {\sl Phys. Rev. Lett.} {\bf 47}(1981)1.
J. Frolich {\sl Nucl. Phys.} {\bf B200}(1982)281.

\bibitem{FREED}
B. Freedman, P. Smolensky, D. Weingarten
{\sl Phys. Lett.} {\bf B113}(1987)25.

\bibitem{CALLPETR}
D.J.E. Callaway and R. Petronzio {\sl Nucl. Phys.} {\bf B267}(1986)253,
{\sl Phys. Lett.}{\bf B139}(1984)189 and {\sl Phys. Lett.} {\bf B145}(1984)381.

\bibitem{O4MODEL}
K. Jansen et al. {\sl Nucl. Phys.} {\bf B265}[FS15](1986)187;

\bibitem{LARKIN}
A.~I.~Larkin, D.~E.~Khmetnontskii. {\sl Sov. Phys. JETP} {\bf 29}(1969)1123.

\bibitem{SHROCK}
R~.~E.~Shrock,{\sl Nucl. Phys} {\bf B267} (1986) 301.

\bibitem{U1GAUGE}
L.A. Fern\'andez, A. Mu\~noz Sudupe, R. Petronzio and A. Taranc\'on,
{\sl Phys. Lett.} {\bf B267} (1991)100.

\bibitem{ALFONSO}
A.~Taranc\'on, {\sl Phys. Rev.} {\bf D36}(1987)3211.

\bibitem{FRADKIN}
E. Fradkin and S.H. Shenker {\sl Phys. Rev} {\bf D19}(1979)3682

\bibitem{VICENTE}
V. Azcoiti, G. di Carlo and A. Grillo, {\sl Phys. Lett.} {\bf B258}(1991)207,
{\bf B268}(1991)101.

\bibitem{STANLEY}
H.E.Stanley. {\it Introduction to Phase Transition and Critical Phenomena}.
Oxford University Press. 1971.

\bibitem{MA}
S.~K.~Ma. {\it Modern Theory of Critical Phenomena}. Benjamin
Inc. 1976.

\bibitem{TRANSPUTERS}
{\it The Transputer databook}. INMOS Databook Series. 1988.

\bibitem{FALCIONI}
M.~Falcioni, E.~Marinari, M.~L.~Paciello, G.~Parisi and
B.~Taglienti, {\sl Phys. Lett.} {\bf B108}(1982)331;
A.~M.~Ferrenberg and R.~Swendsen, {\sl Phys. Rev. Lett.}
{\bf 61}(1988)2635.

\bibitem{FERRENBERG}
A.~M.~Ferrenberg and R.~Swendsen, {\sl Phys. Rev. Lett.}
{\bf 63}(1989)1195.

\bibitem{LETTER}
J.~L.~Alonso et al. {\it The Confining-Higgs phase transition in U(1)-Higgs
LGT}. Preprint FT-UCM 9/92. To be published in {\sl Physics Letters} {\bf B}

\end{thebibliography}
\end{document}